\documentclass[11pt]{article}

\usepackage{amsfonts}
\usepackage{graphicx}
\usepackage{epstopdf}
\usepackage{epsfig}
\usepackage{amssymb}
\usepackage{setspace}
\usepackage{caption}
\usepackage{subcaption}
\usepackage{amsfonts}
\usepackage{color}
\usepackage{amsmath}
\usepackage{float}
\usepackage{comment}
\newcommand {\be}{\begin{equation}}
\newcommand {\ee}{\end{equation}}
\newcommand {\bea}{\begin{array}}
	
	\newcommand {\eea}{\end{array}}

\begin{document}

\begin{titlepage}
	\vspace{1cm} 
	\begin{center}
		{\Large \bf {Accelerating Kaluza-Klein black hole and Kerr/CFT correspondence}}\\
	\end{center}
	\vspace{2cm}
	\begin{center}
		\renewcommand{\thefootnote}{\fnsymbol{footnote}}
		Haryanto M. Siahaan{\footnote{haryanto.siahaan@unpar.ac.id}}\\
		Jurusan Fisika,	Universitas Katolik Parahyangan,\\
		Jalan Ciumbuleuit 94, Bandung 40141, Indonesia
		\renewcommand{\thefootnote}{\arabic{footnote}}
	\end{center}

	\begin{abstract}
	
	We construct a new solution in the Einstein-Maxwell-dilaton theory describing accelerating, charged, and rotating black hole, i.e. the accelerating Kaluza-Klein black hole. Some properties the spacetime are discussed, such as the electromagnetic fields, the area-temperature product, and the holography according to Kerr/CFT correspondence. As expected, the macroscopic Bekenstein-Hawking entropy for an extremal accelerating Kaluza-Klein black hole can be recovered by using Cardy formula of a two dimensional conformal field theory. An interesting feature is found, namely the area-temperature product is just the the one belongs to the vacuum Einstein seed solution.
	
\end{abstract}

\end{titlepage}
	
	\section{Introduction}
	\label{sec:intro}
	
	In vacuum Einstein or electrovacuum systems, there exists a distinctive family of black hole solutions known as the C-metric \cite{Griffiths:2006tk,Kinnersley:1970zw}. These solutions are capable of describing accelerating black holes and are affiliated with the well-established Plebański-Demiański spacetime \cite{Plebanski:1976gy,Griffiths:2005qp}. Numerous studies have not only dedicated into the thermodynamical aspects of the C-metric \cite{Appels:2016uha,Appels:2017xoe,Astorino:2016ybm,Gregory:2017ogk} but have also explored the stability, the superradiance effect, and the validity of strong cosmic censorship in these spacetimes \cite{Destounis:2020pjk,Destounis:2020yav,Destounis:2022rpk,Fontana:2022whx}. In the works of Refs. \cite{Grenzebach:2015oea,Zhang:2020xub}, the authors focused on the theoretical analysis of the shadows cast by accelerating Kerr black holes. Quasinormal modes and overcharging process for the accelerating black holes are studied in \cite{Chen:2024rov} and \cite{Jiang:2023xca}. The thermodynamics of accelerating AdS black hole is investigated in \cite{Kim:2024dbj}, whereas the Schwinger effect for accelerating charged black hole is presented in \cite{Siahaan:2024euo}. 
	
	Among the alternative theories of gravity, the one formulated by Kaluza and Klein in the early 20th century boasts intriguing characteristics. Initially proposed as a classical field theory, it lends itself to interpretations within the realms of quantum mechanics and string theory. Despite experimental searches for the effects of a compact fifth dimension at the Large Hadron Collider~\cite{CMS:2010oej}, no success has been achieved thus far. Further investigations on the aspects of Kaluza-Klein theory have been conducted, i.e. by analyzing its equations of motion and applying it to galactic motion/rotation curves~\cite{Wesson:1992bn,Wesson:1994pj}. Notably, the shadow of a black hole solution within this theory has been scrutinized in~\cite{Amarilla:2013sj}. On another front, forthcoming enhancements in gravitational wave observations might yield constraints on the theory~\cite{Cardoso:2019vof,Andriot:2017oaz,LIGOScientific:2016lio}. In the works of \cite{Azreg-Ainou:2019ylk,Zhu:2020cfn}, the authors investigated the effects of the Kaluza-Klein theory on gyroscope precession and explored potential tests using X-ray spectroscopy. 
	
	Several solutions within the Kaluza-Klein framework have been derived, describing compact objects, notably black holes~\cite{Horowitz:2011cq}. References~\cite{Dobiasch:1981vh,Chodos:1980df,Gibbons:1985ac} present diverse spherically symmetric solutions within the theory. Axially symmetric solutions for both four- and five-dimensional spacetimes have been articulated in Refs.\cite{Larsen:1999pp,Rasheed:1995zv,Matos:1996km}. The investigation of a black hole solution with a squashed horizon is documented in\cite{Ishihara:2005dp,Wang:2006nw}. Solutions for black holes in higher-dimensional spacetimes have been explored in~\cite{Park:1995wk}. The Kaluza-Klein-Taub-NUT spacetime and some of its aspects are studied in \cite{Aliev:2008wv}.

	The Kerr/CFT correspondence \cite{Guica:2008mu} proposes that within the vicinity of the event horizon of an extremal Kerr black hole, under appropriate boundary conditions, there exists a holographic duality with a two-dimensional chiral conformal field theory. This correspondence suggests that the central charge of the conformal field theory is directly proportional to the angular momentum. Furthermore, it has been demonstrated that the macroscopic Bekenstein-Hawking entropy of an extremal Kerr black hole can be reproduced by the microscopic entropy calculation using the Cardy formula of the dual conformal field theory. For an accelerating black hole in Einstein-Maxwell theory, Kerr/CFT correspondence has been discussed in \cite{Astorino:2016xiy}. 
	
	In this paper, we aim to obtain the accelerating version of Kaluza-Klein black hole solution. We use the boosting method \cite{Aliev:2008wv}, with some new approach where we present the new solution in terms of a seed metric. Some aspects of the new solution are studied, namely the electromagnetic field and area-temperature product. In our examination of the electromagnetic field, we draw comparisons with results obtained for the accelerating Kerr-Newman (AKN) case. We also work out the holography for accelerating Kaluza-Klein (AKK) black holes following the Kerr/CFT correspondence introduced in \cite{Guica:2008mu}. 
	
	Organization of this paper is as follows. In the next section, we construct the AKK black hole solution. Section \ref{sec.EM} discusses the electromagnetic field properties in the spacetime, and provides some comparison to the AKN case. The area-temperature product investigation is given in section \ref{sec.AT}, whereas the microscopic entropy calculation is given in section \ref{sec.KerrCFT}. Finally, we have a conclusion. In this paper, we adopt natural units where $c$, $G$, and $\hbar$ are all set to unity.

\section{Solution construction}\label{sec.AKK}

The exposition in this section closely follows the framework outlined in reference \cite{Aliev:2008wv}, incorporating novel methodologies for expressing the transformed solutions in terms of the seed metric. The effective action for Kaluza-Klein theory can be written as \cite{Aliev:2008wv}
\be 
S = \int {d^4 x\sqrt { - g} \left[ {R - 2\partial _\mu  \Phi \partial ^\mu  \Phi  - e^{2\sqrt 3 \Phi } F_{\mu \nu } F^{\mu \nu } } \right]} 
\ee
where $R$ is the four dimensional Ricci scalar, $\Phi$ is the dilation field, $F_{\mu\nu} = \partial_\mu A_\nu - \partial_\nu A_\mu$ is the field-strength tensor related to the $U(1)$ gauge field $A_\mu$. The equations of motion related to this action after varying with respect to each field contents are
\be\label{eq.Einstein} 
R_{\mu \nu }  = 2\partial _\mu  \Phi \partial _\nu  \Phi  + 2e^{2\sqrt 3 \Phi } \left[ {F_{\mu \alpha } F_\nu ^\alpha   - \frac{1}{4}g_{\mu \nu } F_{\alpha \beta } F^{\alpha \beta } } \right]\,,
\ee 
\be \label{eq.Maxwell}
\nabla _\mu  \left( {e^{2\sqrt 3 \Phi } F^{\mu \nu } } \right) = 0\,,
\ee 
and
\be \label{eq.Phi}
2\nabla ^2 \Phi  - \sqrt 3 e^{2\sqrt 3 \Phi } F_{\alpha \beta } F^{\alpha \beta }  = 0\,.
\ee 
There exists a method in constructing a solution obeying the equations of motion above by using a known four dimensional vacuum Einstein system as a seed. First, we uplift a four dimensional vacuum Einstein metric with the metric ${\tilde g}_{\mu\nu}$ to five dimensions as the following
\be 
ds_5^2  = \tilde g_{\mu \nu } dx^\mu  dx^\nu   + dz^2  = G_{MN} dx^M dx^N \,,
\ee 
where we used the coordinate $x^\mu = \left\{t,r,x=\cos\theta,\phi \right\}$ and $x^M = \left\{ x^\mu,z \right\}$. Clearly, the five dimensional metric above solve the five dimensional vacuum Einstein equation ${R}_{M N}=0$, and each component of four dimensional Ricci tensor associated to the metric ${\tilde g}_{\mu\nu}$ also vanishes. 

To get a new solution in Kaluza-Klein theory, the following map known as boosting transformation can be performed \cite{Aliev:2008wv}
\be \label{eq.boost}
\begin{array}{l}
	dt \to \cosh \alpha~dt + \sinh \alpha ~dz\,, \\ 
	dz \to \cosh \alpha~dz + \sinh \alpha ~dt\,, \\ 
\end{array}
\ee 
where the boost velocity is given by $v = \tanh \alpha$. This yields the transformed metric to read
\be \label{eq.metric5dboosted}
ds_5^2  = H^{ - 1} g_{\mu \nu } dx^\mu  dx^\nu   + H^2 \left( {dz + 2A_\alpha  dx^\alpha  } \right)^2 \,,
\ee 
where 
\be 
H^2 = \cosh^2 \alpha + {\tilde g}_{tt} \sinh^2 \alpha \,.
\ee 
In a tetrad form, the four dimensional $ds^2 = g_{\mu\nu}dx^\mu dx^\nu$ can be expressed as
\be 
ds^2  = \eta _{\left( i \right)\left( j \right)} {\pmb \omega} ^{\left( i \right)} {\pmb \omega} ^{\left( j \right)} \,,
\ee 
where the corresponding one-form is given by
\[
{\pmb \omega} ^{\left( 0 \right)}  = \sqrt { \frac{{\tilde g}_{tt}}{H} } \left( {dt + c \frac{{\tilde g}_{t\phi}}{{\tilde g}_{tt}}d\phi } \right)
~,~
{\pmb \omega} ^{\left( 1 \right)}  = \sqrt{H {\tilde g}_{rr}}~ dr
~,~
{\pmb \omega} ^{\left( 2 \right)}  = \sqrt{H {\tilde g}_{xx}}~ dx
~,~
{\pmb \omega} ^{\left( 3 \right)}  = \sqrt { \frac{H}{{\tilde g}_{tt}} \left( {\tilde g}_{\phi\phi}{\tilde g}_{tt} - {\tilde g}_{t\phi}^2 \right)} d\phi
\]
and the non-vanishing components of a flat Minkowski metric are $\eta _{\left( 1 \right)\left( 1 \right)}=\eta _{\left( 2 \right)\left( 2 \right)}=\eta _{\left( 3 \right)\left( 3 \right)}=-\eta _{\left( 0 \right)\left( 0 \right)}=1$. The set of field contents $\left\{g_{\mu\nu}, A_\mu, \Phi \right\}$ that solves the equations of motion (\ref{eq.Einstein}) - (\ref{eq.Phi}) consists of the four dimensional metric tensor $g_{\mu\nu}$ in eq. (\ref{eq.metric5dboosted}), the gauge vector
\be 
A_\mu  dx^\mu   = \frac{{\sinh \alpha \left[ {\left( {1 + \tilde g_{tt} } \right)\cosh \alpha dt + \tilde g_{t\phi } d\phi } \right]}}{{2H^2 }}\,,
\ee
and the dilaton field  
\be 
\Phi = \frac{\sqrt{3}}{2} \ln H \,.
\ee 
and 

Now let us employ the prescription above to the accelerating Kerr metric 
\be \label{eq.metricAccKerr}
ds^2  = \frac{1}{{\Omega ^2 }}\left[ -{\frac{{\Delta _r }}{\Sigma }\left( {dt - a\Delta _x d\phi } \right)^2  + \Sigma \left( {\frac{{dr^2 }}{{\Delta _r }} + \frac{{dx^2 }}{{P\Delta _x }}} \right) + \frac{{P\Delta _x }}{\Sigma }\left( {adt - \left( {r^2  + a^2 } \right)d\phi } \right)^2 } \right]
\ee 
as the ${\tilde g}_{\mu\nu}$. In equation above, $r_\pm = m \pm \sqrt{m^2 -a^2}$, $P = \left( {1 - bxr_ +  } \right)\left( {1 - bxr_ -  } \right)$, $\Omega = 1-brx$, $\Sigma = r^2 + a^2 x^2$, $\Delta_x =1-x^2$, and
\be 
\Delta _r  = \left( {r - r_ +  } \right)\left( {r - r_ -  } \right)\left( {1 - b^2 r^2 } \right)\,.
\ee 
The parameters involved are $b$ as the acceleration parameter of the black hole, $a$ as the rotational parameter, and $m$ as the black hole mass.

At the poles $x=\pm 1$, we can compute
\[
C_{\pm}    = \mathop {\lim }\limits_{x \to  \pm 1} \frac{1}{{\Delta _x }}\sqrt {\frac{{g_{\phi \phi } }}{{g_{xx} }}}  = 1 \mp 2bm + a^2 b^2 \,.
\]
One can infer from the last equation that the conical defects at the poles cannot be cured by scaling the angular coordinate $\phi$, i.e. $\phi \to \Delta_\phi \phi$ for some constants $\Delta_\phi$. This conical singularity is interpreted as the source of black hole acceleration, in the form of string or strut. However, we can ensure regularity on one axis. Specifically, in this paper, we will apply it to the semi-infinite axis at $x=1$, by scaling $\phi \to  \phi/C_+$. This scaling in the seed solution will be explicitly reflected in the boosted solution presented below. It is important to note that the scaling required to maintain the $2\pi$ periodicity on the $x=1$ axis remains consistent in both the seed and boosted metric solutions.

Using the prescription above, the metric describing accelerating rotating and charged black holes in Kaluza-Klein theory can be written as
\[
ds^2  =  - \frac{{P\Delta _x a^2  - \Delta _r }}{{\Sigma \Omega ^2 H_s }}\left( {dt + \frac{{a\Delta _x \left( {\Delta _r  - P\left( {r^2  + a^2 } \right)} \right)}}{{\left(P\Delta _x a^2  - \Delta _r\right) C_+ }}d\phi } \right)^2 
\]
\be\label{eq.metricKK}
+ \frac{{H_s \Sigma }}{{\Omega ^2 }}\left( {\frac{{dr^2 }}{{\Delta _r }} + \frac{{dx^2 }}{{P\Delta _x }}} \right) + \frac{{H_s \Delta _x \Delta _r P\Sigma }}{{\Omega ^2 \left( {\Delta _r  - P\Delta _x a^2 } \right) C_+^2}}d\phi ^2  \,,
\ee 
where
\be 
H_s  = \left(\frac{{\left( {P\Delta _x a^2  - \Delta _r } \right)\sinh^2 \alpha + \cosh^2 \alpha \Sigma \Omega ^2 }}{{\Sigma \Omega ^2 }}\right)^{\tfrac{1}{2}}\,.
\ee 
Accordingly, the dilaton field reads
\be 
\Phi  = \frac{{\sqrt 3 }}{4}\ln \left[ {\frac{{\left( {P\Delta _x a^2  - \Delta _r } \right)\sinh^2 \alpha + \cosh^2 \alpha \Sigma \Omega ^2 }}{{\Sigma \Omega ^2 }}} \right]\,,
\ee 
whereas the non-zero component of the gauge field can be written as
\be 
A_t  = \frac{{{\cosh \alpha}{\sinh \alpha} \left( {P\Delta _x a^2  - \Delta _r  + \Sigma \Omega ^2 } \right)}}{{2\left( {\left( {P\Delta _x a^2  - \Delta _r } \right)\sinh^2 \alpha + \cosh^2 \alpha \Sigma \Omega ^2 } \right)}}\,,
\ee 
and
\be 
A_\phi   = \frac{{a\Delta _x {\sinh \alpha} \left( {\Delta _r  - P\left( {r^2  + a^2 } \right)} \right)}}{{2C_+ \left( {\left( {P\Delta _x a^2  - \Delta _r } \right)\sinh^2 \alpha + \cosh^2 \alpha \Sigma \Omega ^2 } \right)}}\,.
\ee

The horizons in AKK spacetime are given by the roots of $\Delta_r$. Similar to its Einstein-Maxwell counterpart, namely AKN background, there exist four horizons; the inner and outer black hole horizons given by $r_\pm = m \pm \sqrt{m^2-a^2}$, and the acceleration horizons $r_\pm^a = \pm b^{-1}$. We limit the discussions for black hole horizons only since the area of horizon becomes singular at $r^a$. The angular velocity at the horizon for the AKK spacetime can be computed as
\be \label{eq.OmH}
\Omega _H  =  - \left. {\frac{{g_{t\phi } }}{{g_{\phi \phi } }}} \right|_{r = r_+ }  = \frac{{C_ +  a}}{{\cosh \alpha \left( {r_+^2  + a^2 } \right)}}
\ee

The parameter $\cosh\alpha$ in the solution generating method above contributes to the conserved quantities in the spacetime. In the non-accelerating case, the conserved Kaluza-Klein black hole mass can be found as \cite{Aliev:2008wv}
\be 
M = \frac{m}{2} \left(1+\cosh^2\alpha\right)\,,
\ee 
by using Komar integral. On the other hand, the black hole angular momentum is
\be 
J = am \cosh\alpha\,.
\ee 
For the identity transformation in (\ref{eq.boost}) denoted by $\cosh\alpha=1$, the transformed metric (\ref{eq.metricKK}) reduces to accelerating Kerr solution of the vacuum Einstein, whereas the non-gravitational fields $\left\{A_\mu,\Phi\right\}$ vanish.

To complete the discussion of new solution presented in this section, let us investigate some of its properties. First, the curvature singularities that can be learned from the squared of Riemann tensor. Indeed, presenting the full terms of this quantity in the general form will be lengthy and not too informative. Therefore, we rather to present the equatorial one and compared it to the seed solution counterpart to understand how the boosting transformation (\ref{eq.boost}) modifies the structure of curvature singularity. In addition, to let the discussion as simple as possible, we also consider the non-rotating case just to examine how the singularity structure gets modification due to the boosting transformation. At equator, the squared of Riemann tensor that corresponds to the metric in eq. (\ref{eq.metricKK}) in $a\to 0$ limit can be expressed as
\be \label{eq.RRx0}
\left. {R_{\alpha \beta \mu \nu } R^{\alpha \beta \mu \nu } } \right|_{x = 0}  = \frac{{c_0  + c_2 \cosh ^2 \alpha  + c_4 \cosh ^4 \alpha  + c_6 \cosh ^6 \alpha  + c_8 \cosh ^8 \alpha }}{{4r^5 \left\{ {\left( {2m - r} \right)\left( {1 - b^2 r^2 } \right) - \cosh ^2 \alpha \left( {2m\left( {1 - b^2 r^2 } \right) + b^2 r^3 } \right)} \right\}}}
\ee
where the functions $c_i$'s are given in the appendix. In $\cosh\alpha=1$ case, namely the seed solution (\ref{eq.metricAccKerr}), the squared Riemann tensor above reduces to 
\be \label{eq.RRcmetric}
\left. {R_{\alpha \beta \mu \nu } R^{\alpha \beta \mu \nu } } \right|_{x = 0}  = \frac{{48\left( {1 - brx} \right)^6 m^2 }}{{r^6 }}
\ee 
which belongs to the C-metric. Comparing the last two equations tells us that boosting transformation (\ref{eq.boost}) does modify the singularity structure of the spacetime. From the true singularity at the origin for the seed solution as reflected in (\ref{eq.RRcmetric}) becomes the locations of singularities that are given by the roots of
\be \label{eq.singularityEX}
{\left( {2m - r} \right)\left( {1 - b^2 r^2 } \right) - \cosh ^2 \alpha \left( {2m\left( {1 - b^2 r^2 } \right) + b^2 r^3 } \right)} = 0
\ee 
in addition to the generic one at $r=0$. This situation perhaps appears to be strange, considering the possibility for existence of physical singularity outside the horizon. Note that, even in the non-accelerating case of Kaluza-Klein black hole, such peculiarity appears as suggested by eq. (\ref{eq.singularityEX}) in the $b\to 0$ limit. 

\begin{figure}[H]
	\begin{center}
		\includegraphics*[scale=0.35]{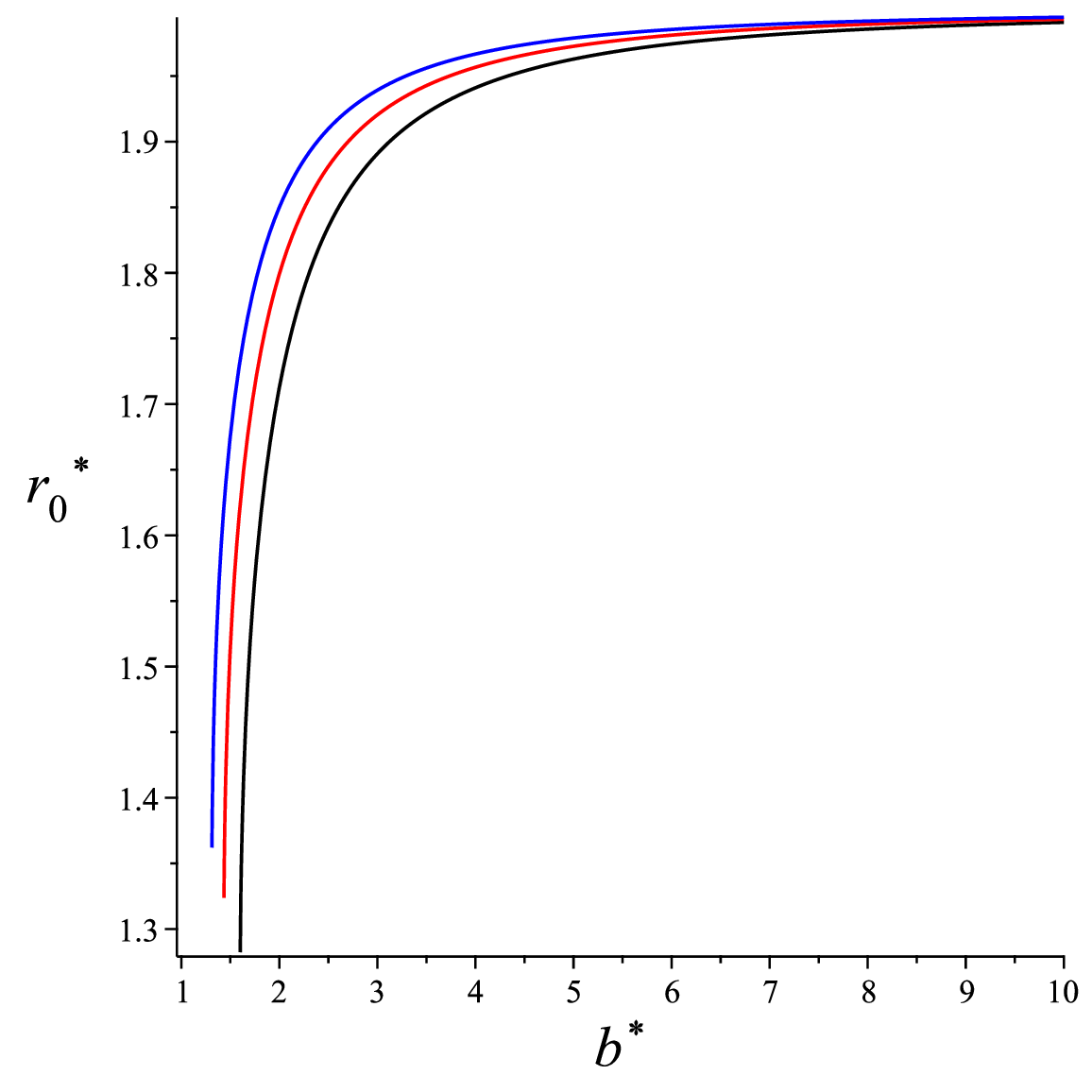}\caption{Some numerical evaluations of $r_0$ according to (\ref{eq.r0}). The black, red, and blue curves represent the cases of $\cosh^2\alpha=1.5$, $\cosh^2\alpha=2$, $\cosh^2\alpha=6$, respectively.}\label{fig.rc}
	\end{center}
\end{figure}

However, a careful analysis can be performed to show that such extra spacetime singularities are still located inside the event horizon, for the non-rotating case. By using some symbolic manipulation programs such as MAPLE, the qubic equation (\ref{eq.singularityEX}) can be solved and yields the solution
\be \label{eq.r0}
r_0^*  = \frac{{\left( {{\cal Z}\sinh ^4 \alpha } \right)^{1/3} }}{{3b^* \sinh ^2 \alpha }} + \frac{{4b^{*2} \sinh ^2 \alpha  - 3}}{{3b^* \left( {{\cal Z}\sinh ^4 \alpha } \right)^{1/3} }} + \frac{2}{3}
\ee 
where
\[
{\cal Z} = 
8b^{*3} \sinh ^2 \alpha  - 27b^* \cosh ^2 \alpha  + 18b^*  + \frac{3}{{\sin \alpha }}\left[ 1-16 b^{*4} \cosh^{6}{\alpha} +48 b^{*4} \cosh^{4}{\alpha} +27 b^{*2} \cosh^{6}{\alpha}\right.
\]
\[
\left.  -48 b^{*4} \cosh^{2}{\alpha}-63 b^{*2} \cosh^{4}{\alpha}+16 b^{*4}+44 \cosh^{2}{\alpha} b^{*2}-8 b^{*2} \right]\,.
\]
Note that we have presented $r_0^*$ as a dimensionless quantity and $b^* = bm$. Interestingly, independent of the value $\cosh\alpha > 1$ under consideration, one can compute that
\be 
\mathop {\lim }\limits_{b \to \infty } r_0^*  = 2\,.
\ee 
It suggests the radius that leads to the singular value of the squared Riemann tensor (\ref{eq.RRx0}) apart from the origin is kept to be hidden inside the horizon for some finite acceleration parameter $b$. Plots in fig. \ref{fig.rc} illustrate this behavior, namely the asymptotic value of $r_0^*$ for large $b$.

\begin{figure}[H]
	\begin{center}
		\includegraphics*[scale=0.4]{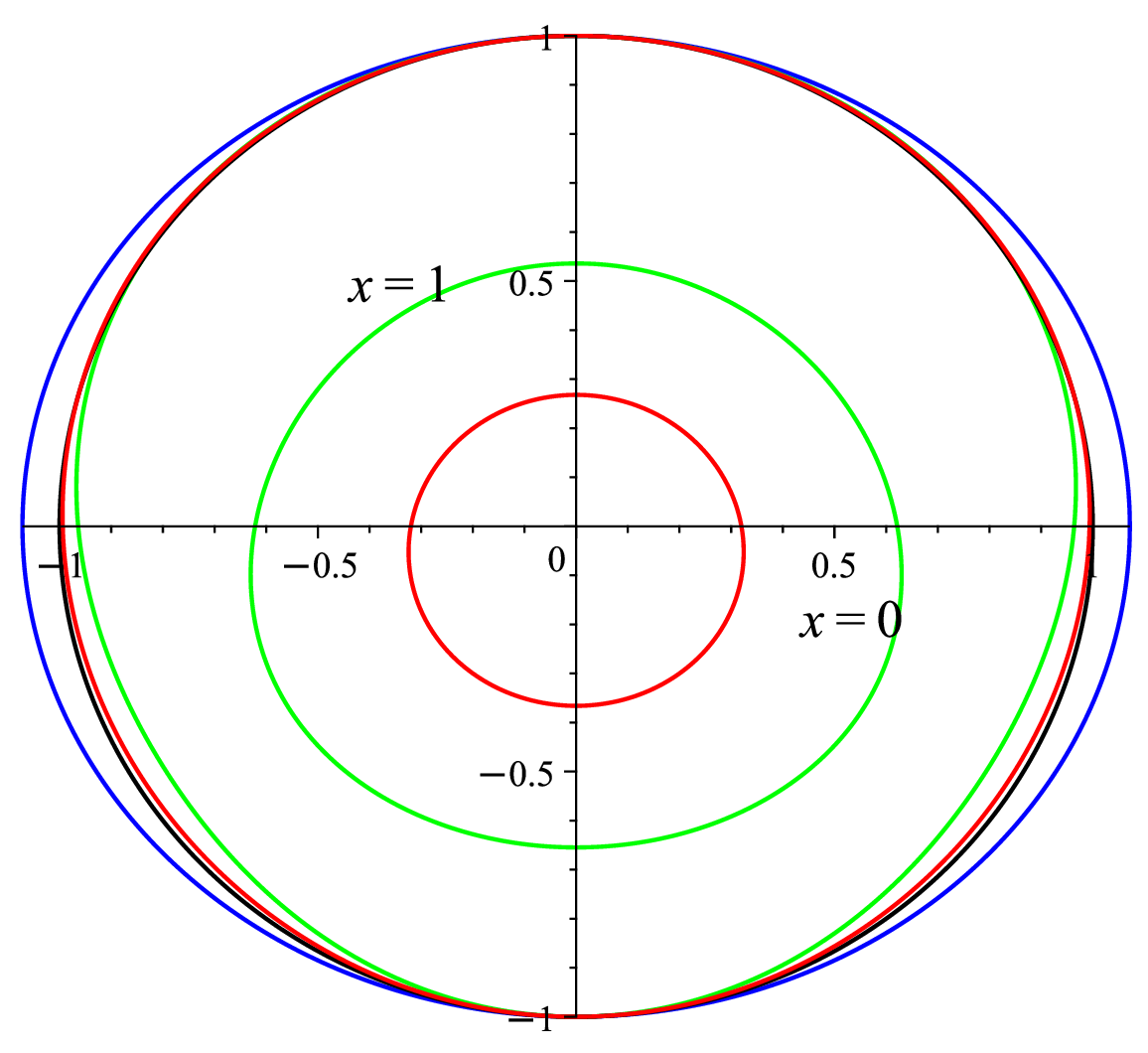}\caption{Numerical evaluations of some ergoregions boundaries in the accelerating Kaluza-Klein black hole spacetime. The blue cure represents the boundary of ergoregion for the non-accelerating case, black curve denotes the black hole outer horizon, whereas the green and red curves describe the cases of $bm=1$ and $bm=2$, respectively.}\label{fig.ergo}
	\end{center}
\end{figure}

Now let us turn to the ergoregions discussion. It is known that $g_{tt}=0$ denotes the boundary of ergoregion. It turns out that the condition that determines the ergoregion boundary for accelerating Kaluza-Klein black hole is identical to that of the seed solution (\ref{eq.metricAccKerr}), namely
\be \label{eq.Ergo}
\Delta \left( {r = r_e } \right) = P\Delta _x a^2 \,.
\ee 
Therefore, the radius of ergoregion boundary $r_e$ depends on the black hole parameters $a$, $b$, $m$, and the angular coordinate $x=\cos\theta$ as well. A thorough analysis of ergoregion in an accelerating spacetime even with NUT parameter has been performed in \cite{Podolsky:2021zwr}. 
Let us revisit some relevant plots to examine how the acceleration of the black hole influences the boundary of the ergoregion. As shown in figure \ref{fig.ergo}, acceleration plays a significant role in altering these boundaries. The larger red and green curves in the figure display an asymmetry between the $x>0$ and $x<0$ regions. This contrasts with the Kerr spacetime, where, although the ergoregion is not spherical, the radius of the ergoregion remains symmetric under the coordinate transformation $x \to -x$.

\begin{figure}[H]
	\begin{center}
		\includegraphics*[scale=0.4]{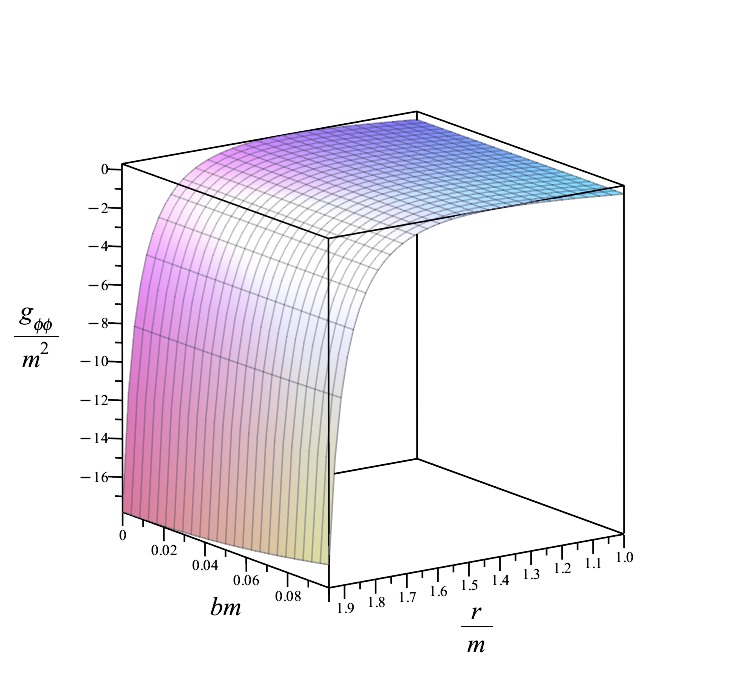}\caption{Numerical evaluations of $g_{\phi\phi}$ where we consider $m=1$, $a=0.5$, and $\cosh\alpha=1.1$.}\label{fig.ctc}
	\end{center}
\end{figure}

Additionally, we can explore the presence of closed timelike curves (CTCs) within this spacetime. Given that we are using the mostly positive convention for the spacetime signature, the existence of a CTC is indicated by negative values of $g_{\phi\phi}$, namely given by the condition
\be \label{eq.CTC1}
\frac{{a^2 \left( {\Delta _r  - P\left( {r^2  + a^2 } \right)} \right)^2 \Delta _x  + H_s^2 \Delta _r P\Sigma ^2 }}{{\left( {\Delta _r  - P\Delta_x a^2 } \right)\Sigma \Omega ^2 H_s }} < 0\,.
\ee 
Clearly, the final inequality is difficult to solve analytically due to the complexity of the functions and the numerous parameters involved. However, to demonstrate the existence of CTCs, we present a numerical result in fig. \ref{fig.ctc} indicating that for certain chosen parameter values, $g_{\phi\phi}$ can indeed take on negative values.

\section{Electromagnetic fields}\label{sec.EM}

In this section, we will explore various aspects of the electromagnetic field in both AKN and AKK spacetimes and make some comparisons. Note that some aspects of AKN spacetime have been discussed in \cite{Sui:2023rfh,BarraganAmado:2023wxt,Kraniotis:2021qah,Chen:2021tbo,Podolsky:2006px}. The AKN solution can be expressed by using exactly the same form as in eq. (\ref{eq.metricAccKerr}) with the only modification for inner and outer black hole horizon radius,
\be \label{eq.rpmAccKN}
r_\pm = m \pm \sqrt{m^2 - a^2 -q^2}\,.
\ee 
As an electrovacuum system, the metric (\ref{eq.metricAccKerr}) with horizons (\ref{eq.rpmAccKN}) together with the gauge field
\be 
A_\mu dx^\mu = \frac{qr}{\Sigma} \left(dt-a \Delta_x d\phi \right)
\ee 
solve the Einstein-Maxwell equations 
\be 
R_{\mu\nu} = 2 F_{\mu \kappa} F_{\nu }^\kappa - \frac{1}{2} g_{\mu\nu} F_{\alpha\beta}F^{\alpha\beta}
\ee 
where $q$ is the black hole charge. The general form of electric and magnetic field experienced by an observer with four-velocity $u^\mu$ are given by
\be 
E_\mu = F_{\mu \nu} u^\nu \,,
\ee 
and
\be 
B_\mu = \frac{1}{2} \epsilon_{\mu\nu\alpha\beta} F^{\alpha\beta} u^\nu \,,
\ee 
respectively. 

In this discussion, let us consider a static observer with four-velocity $u^\mu = [1,0,0,0]$. Interestingly, for this observer, the radial and polar components of the electric field in AKN spacetime do not depend on the acceleration parameter $b$. In other words, the non-vanishing components of the electric fields are just that of non-accelerating counterpart. These fields in AKN spacetime can be written as
\be \label{eq.EAccKN}
E_r  =  - \frac{{q\left( {r^2  - a^2 x^2 } \right)}}{{\left( {r^2  + a^2 x^2 } \right)^2 }}\,\, ,\,\,
E_x  =  - \frac{{2qra^2 x}}{{\left( {r^2  + a^2 x^2 } \right)^2 }}\,,
\ee 
\be \label{eq.BAccKN}
B_r  =  - \frac{{2aqrx\left( {1 - brx} \right)^4 D_+ }}{{\left( {r^2  + a^2 x^2 } \right)^3 }} \,\, ,\,\,
B_x  = \frac{{qa\left( {r^2  - a^2 x^2 } \right)\left( {1 - brx} \right)^4 D_+ }}{{\left( {r^2  + a^2 x^2 } \right)^3 }}\,.
\ee 
From the above expressions, we observe that $B_r$ and $E_x$ vanish at the equator. Above, we have considered the AKN spacetime with regularity on $x=1$ axis, which is ensured by the scale factor $D_+ = 1-2mb +b^2 \left(a^2+q^2\right)$.

Now, let us shift our focus to the Accelerating Kaluza-Klein scenario. Note that the complete expressions for the electric and magnetic fields observed by a stationary observer are rather extensive. Therefore, we will only provide the outcomes for the equatorial and polar regions, allowing a comparison with the AKN case. The non-vanishing equatorial electric and magnetic fields components with respect to an observer with four-velocity $u^\mu$ are
\be \label{eq.Er.accKKx0}
E_r \left( {x = 0} \right) = \frac{{\cosh \alpha \sinh \alpha \left( {m + r^2 b^2 \left( {m - r} \right)} \right)}}{{\left( {\left( {r - r_ +  } \right)\left( {r - r_ -  } \right)\sinh^2\alpha rb^2  + r + 2m\sinh^2\alpha} \right)^2 }}\,,
\ee 
\be \label{eq.Ex.accKKx0}
E_x \left( {x = 0} \right) = \frac{{b\cosh \alpha \sinh \alpha \left( {r^3 \left( {r - r_ +  } \right)\left( {r - r_ -  } \right) + 2mr^2  - ma^2  - r^3 } \right)}}{{\left( {\left( {r - r_ +  } \right)\left( {r - r_ -  } \right)\sinh^2\alpha rb^2  + r + 2m\sinh^2\alpha} \right)^2 }}\,,
\ee 
\be \label{eq.Br.accKKx0}
B_r \left( {x = 0} \right) =  - \frac{{bma\sinh\alpha \left( {r^3 \left( {r - r_ +  } \right)\left( {r - r_ -  } \right) + 2mr^2  - ma^2  - r^3 } \right)}}{{r^2 \left( {\left( {r - r_ +  } \right)\left( {r - r_ -  } \right)\sinh^2\alpha rb^2  + r + 2m\sinh^2\alpha } \right)^2 }}\,,
\ee
and 
\be \label{eq.Bx.accKKx0}
B_x \left( {x = 0} \right) = \frac{{a\sinh \alpha \left( {m + r^2 b^2 \left( {m - r} \right)} \right)}}{{r^2 \left( {\left( {r - r_ +  } \right)\left( {r - r_ -  } \right)\sinh^2\alpha rb^2  + r + 2m\sinh^2\alpha } \right)^2 }}\,.
\ee 
Interestingly, unlike in the case of AKN spacetime, the general expressions for equatorial $E_x$ and $B_r$ in AKK spacetime are not vanished. This marks a distinct difference in the electromagnetic properties between the AKN and AKK spacetimes. 

\begin{figure}[H]
	\begin{center}
		\includegraphics*[scale=0.3]{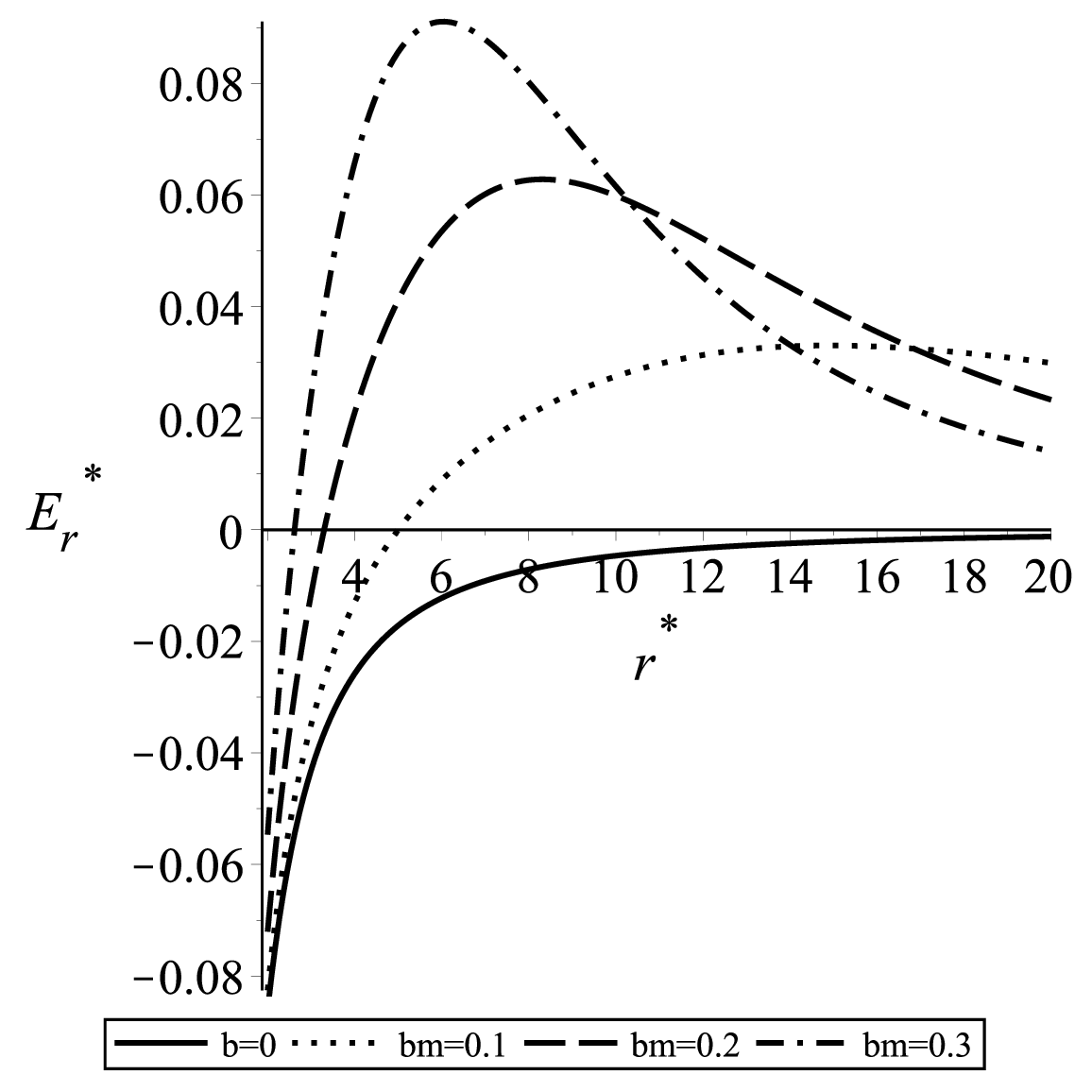}\caption{Equatorial dimensionless $E_r$ in AKK spacetime. In the absence of the acceleration parameter, it is evident that the electric field results in a purely attractive interaction for a positive test electric charge. However, with an increase in the acceleration parameter, the same test charge may experience repulsion from the accelerating black hole. This distinctive feature is absent in the electric field of the AKN spacetime as described by the eq. (\ref{eq.EAccKN}). Note that the corresponding field in AKN spacetime in eq. (\ref{eq.EAccKN}) is independent of $b$.}\label{fig.Erx0KK}
	\end{center}
\end{figure}

\begin{figure}[H]
	\begin{center}
		\includegraphics*[scale=0.3]{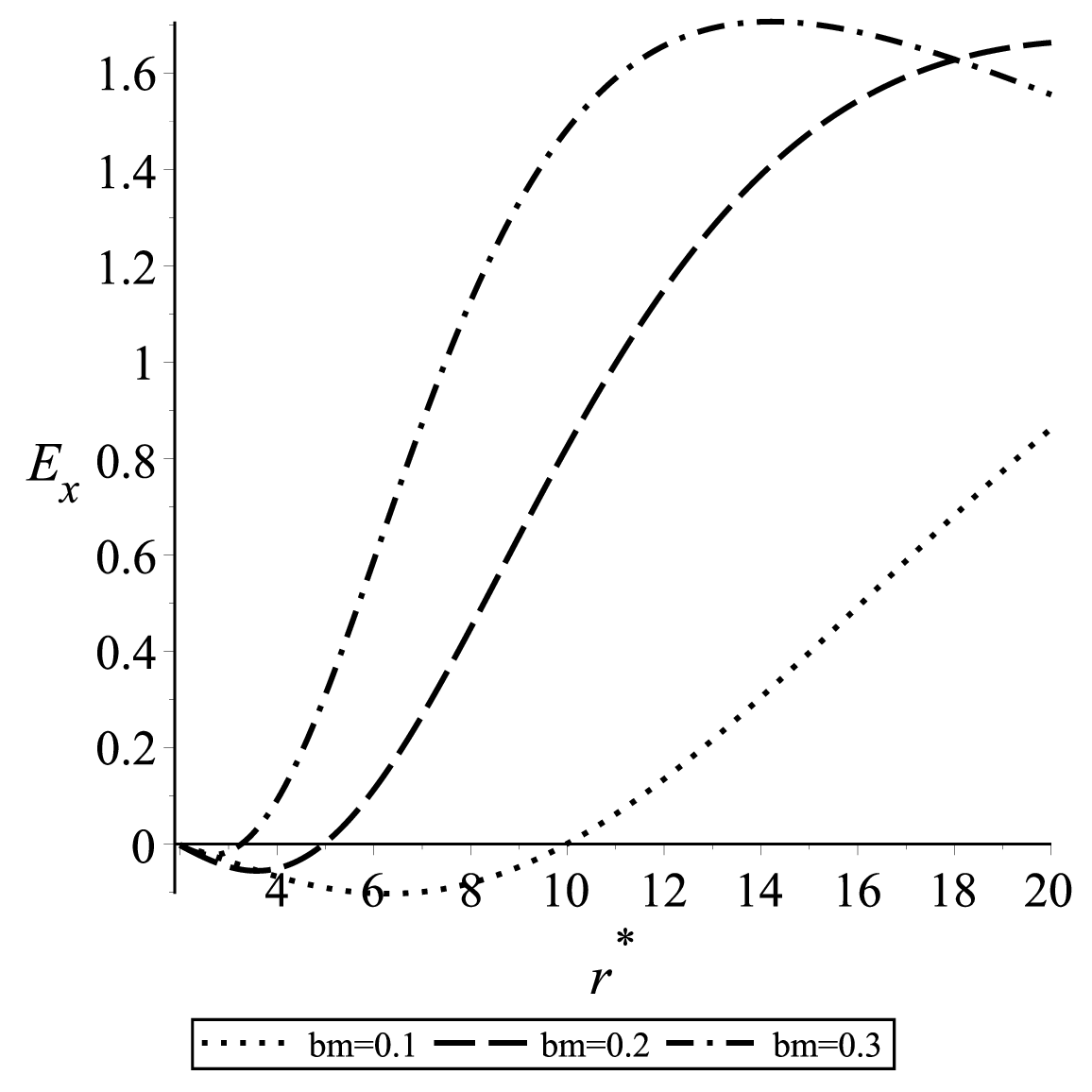}\caption{Equatorial dimensionless $E_x$ in AKK spacetime. The corresponding fields in AKN spacetime is given in eq. (\ref{eq.EAccKN}) and it vanishes at equator.}\label{fig.Exx0KK}
	\end{center}
\end{figure}

\begin{figure}[H]
	\begin{center}
		\includegraphics*[scale=0.3]{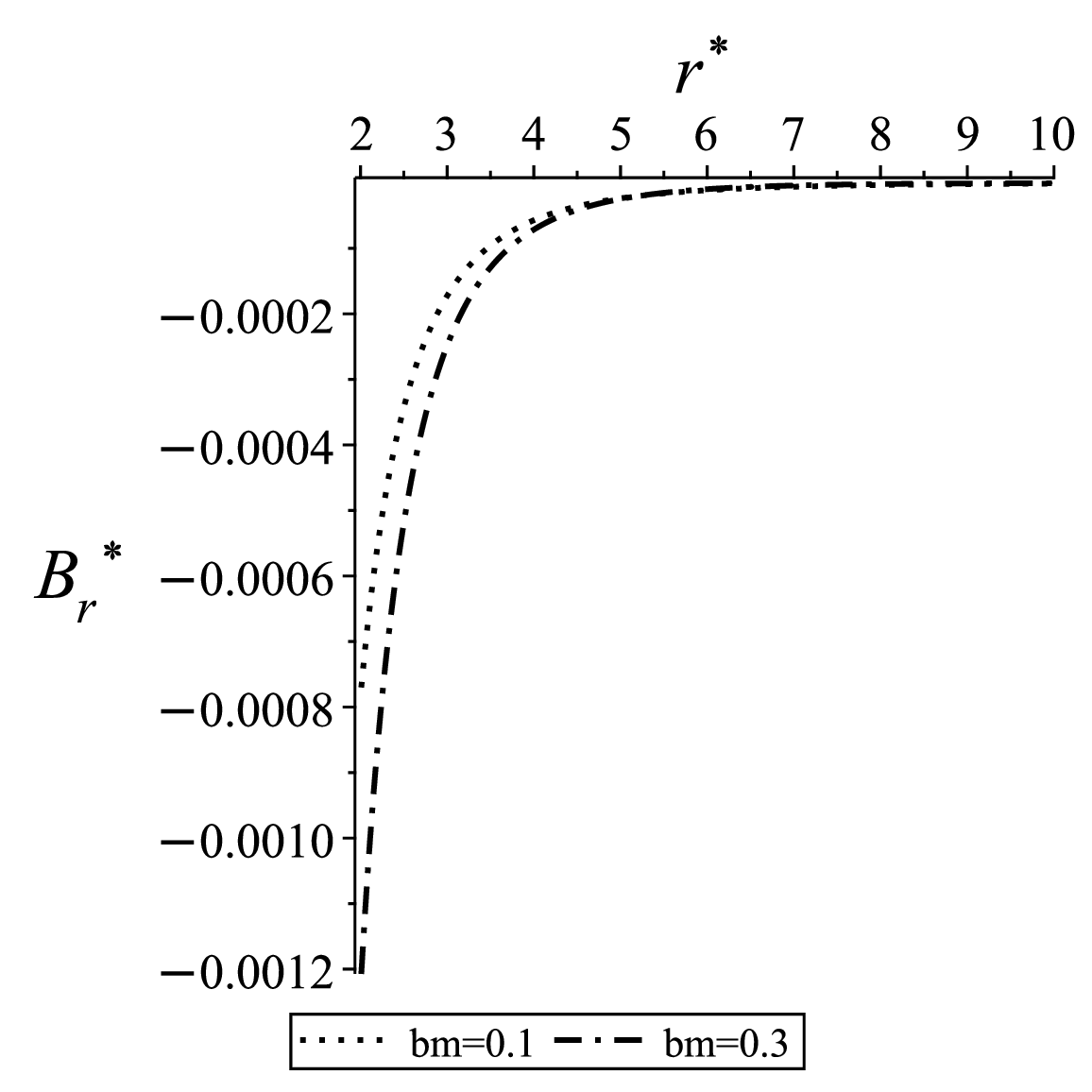}\caption{Equatorial dimensionless $B_r$ in AKK spacetime. The corresponding fields in AKN spacetime is given in eq. (\ref{eq.BAccKN}) which is vanished at equator.}\label{fig.Brx0KK}
	\end{center}
\end{figure}

\begin{figure}[H]
	\begin{center}
		\includegraphics*[scale=0.3]{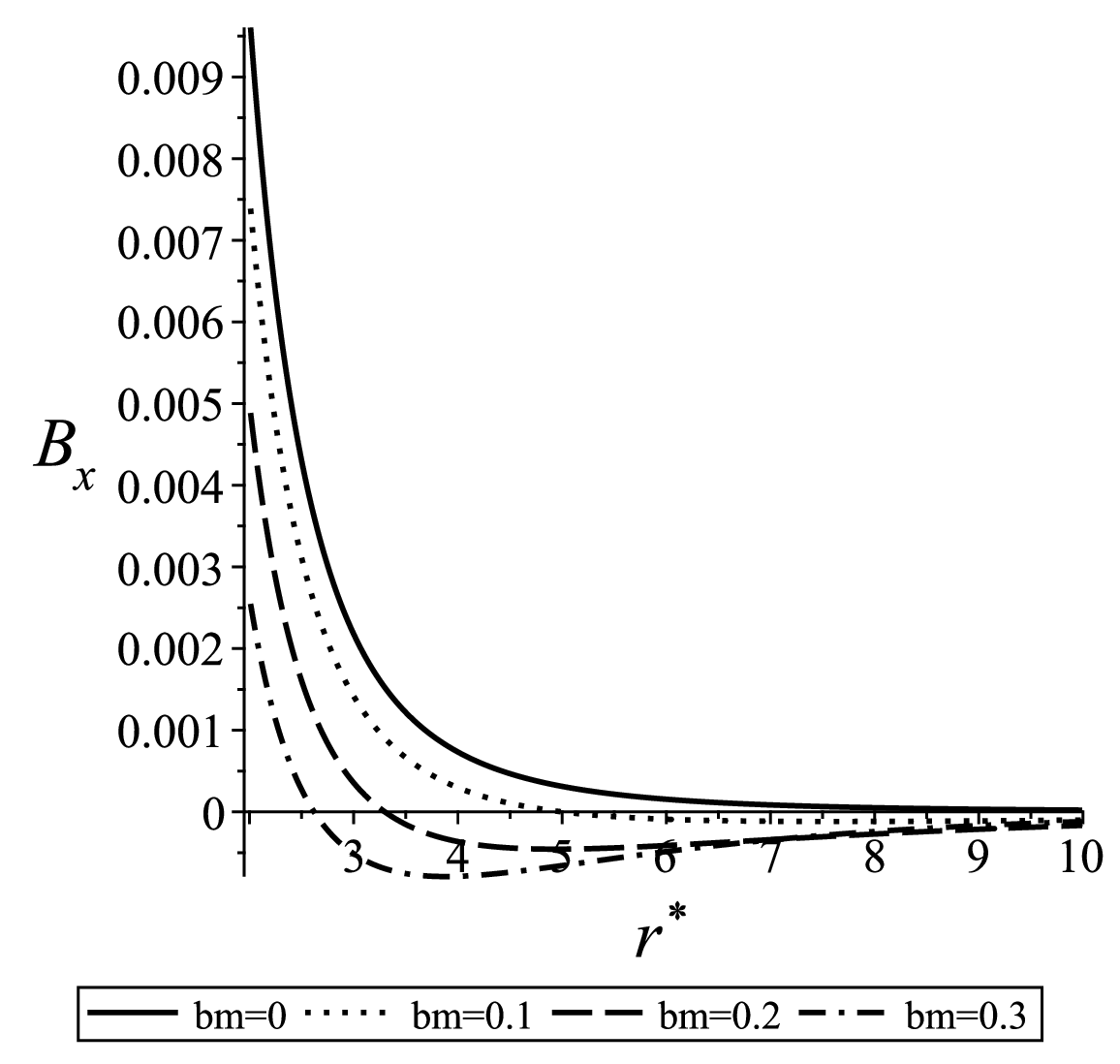}\caption{Equatorial dimensionless $B_x$ in AKK spacetime.}\label{fig.Bxx0KK}
	\end{center}
\end{figure}

Obviously, the functions (\ref{eq.Er.accKKx0})-(\ref{eq.Bx.accKKx0}) above are not too intuitive. Therefore we provide selected numerical evaluations to simplify the comparison with the findings from the AKN scenario. For all the numerical plots below, we consider $a=0.5~m$ and $\cosh \alpha=1.1$. From all the figures \ref{fig.Erx0KK}, \ref{fig.Exx0KK}, \ref{fig.Brx0KK}, and \ref{fig.Bxx0KK}, it can be observed that the equatorial electromagnetic fields differ between the AKK and AKN cases in general.


\begin{figure}[H]
	\begin{center}
		\includegraphics*[scale=0.3]{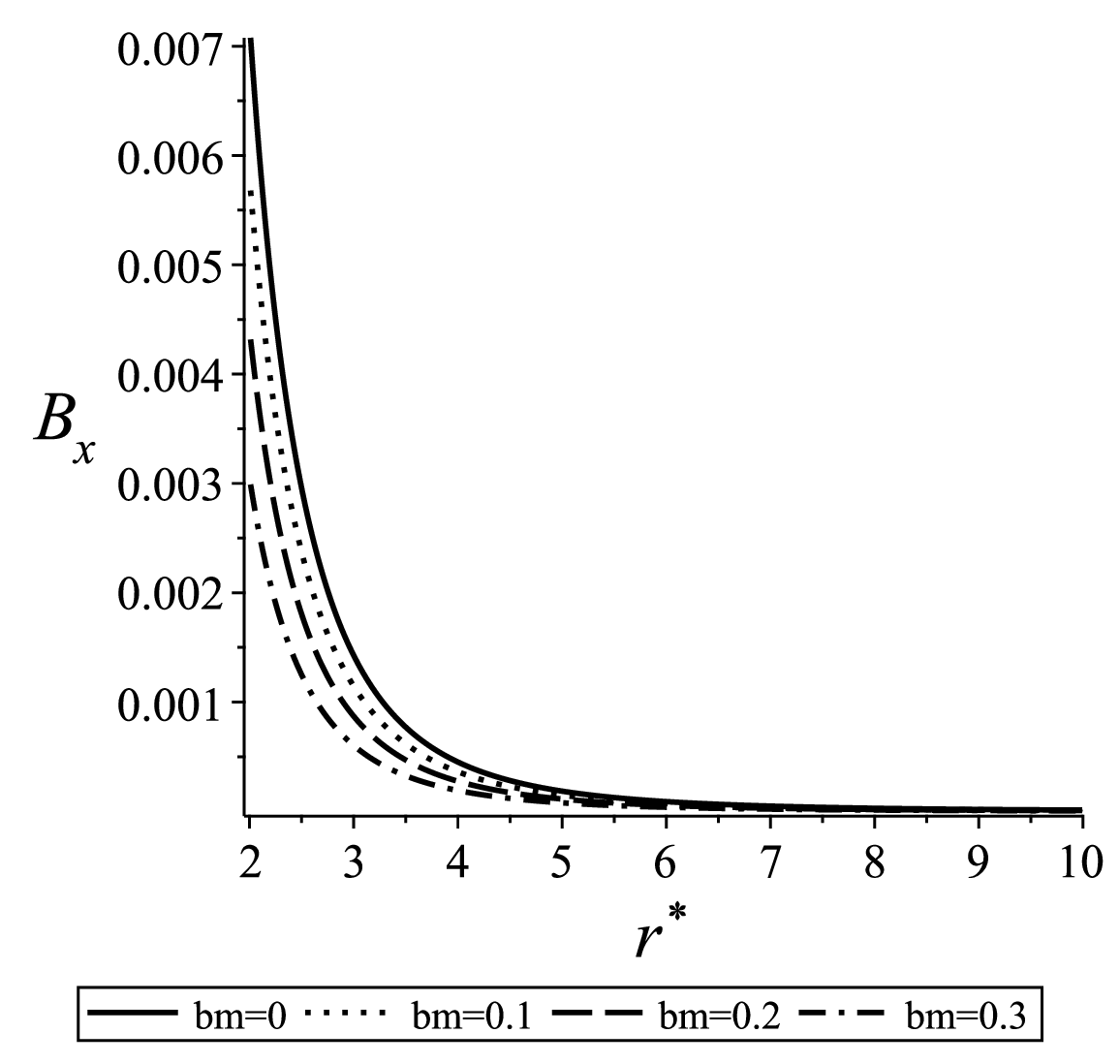}\caption{Equatorial dimensionless $B_x$ in AKN spacetime as given in eq. (\ref{eq.BAccKN}). Here we consider $q=0.5~m$, comparable to the choice in AKK where $\cosh\alpha=1.1$.}\label{fig.Bxx0KN}
	\end{center}
\end{figure}

Another distinct electromagnetic property between the AKN and AKK spacetimes can be observed in Figs. \ref{fig.Bxx0KK} and \ref{fig.Bxx0KN}. For a similar numerical value of the black hole charge, specifically $0.5~m$\footnote{We consider the electric charge in the non-accelerating spacetime for both AKN and AKK, where for AKK it is $q=m\cosh\alpha\sinh\alpha$ \cite{Aliev:2008wv}.}, one can see that the plot of the $x$-component of the magnetic field in the AKK spacetime shows a local minimum outside the event horizon as the acceleration parameter increases. This behavior does not manifest in the AKN spacetime.

\section{Area-Temperature product}\label{sec.AT}

The black hole entropy is determined by the quarter of the area law
\be 
S_h = \frac{{\cal A}_h}{4}\,.
\ee 
Using the standard formula for horizon's area, one can compute for the AKK black hole that
\be 
{\cal A}_h = 
\int\limits_{x =  - 1}^1 {\int\limits_0^{2\pi } {{\left. {\sqrt {g_{xx} g_{\phi \phi } } } \right|_{r=r_h } } dxd\phi } } \,,
\ee 
which leads to
\be \label{eq.area}
{\cal A}_h^{\pm} = \frac{{4\pi \cosh\alpha \left( {r_{\pm}^2  + a^2 } \right)}}{{\left( {1 - b^2 r_\pm^2 } \right) C_+}}\,,
\ee 
after using the component of the metric tensor $g_{\mu\nu}$ as appeared in eq. (\ref{eq.metricKK}). Note that for $\cosh\alpha=1$ which is the case of identity mapping in the boost transformation above, we have the area of accelerating Kerr spacetime horizons. 

Related to the are and temperature of black hole, there exist an interesting relation namely the area-temperature product associated to the outer and inner horizons 
\be \label{eq.areatempPROD}
{\cal A}_{h}^+ T_{H}^+ = - {\cal A}_{h}^- T_{H}^- \,.
\ee 
The authors of \cite{Podolsky:2021zwr} showed that relation (\ref{eq.areatempPROD}) holds for the AKN black hole. To establish such relationship for the AKK case, first we need to compute the Hawking temperature associated with the horizon under consideration. Here we use the tunneling method, where the general formula for temperature is given by \cite{Siahaan:2021ags}
\be 
T_H^{\pm}  = \left. {\frac{{\sqrt { - \left( {\partial _r g_{tt} } \right)\left( {\partial _r g^{rr} } \right)} }}{{4\pi }}} \right|_{r = r_\pm } \,.
\ee 
Using this formula, the Hawking temperature can be read as
\be \label{eq.temp}
T_H^+   = \frac{{m-r_\pm +b^2 \left( {2r_\pm^3  + r_\pm a^2  - 3mr_\pm^2 } \right)}}{{2\pi \cosh\alpha\left( {r_+^2  + a^2 } \right)}}\,.
\ee 

The product of Hawking temperature (\ref{eq.temp}) and the area of black hole horizon given in eq. (\ref{eq.area}) can be expressed as
\be \label{eq.AreaTempProdgen}
{\cal A}_{h}^{\pm} T_{H}^\pm =\pm  \frac{2\left(m-r_{+} +b^2 \left( {2r_{+}^3  + r_{+} a^2  - 3mr_{+}^2 } \right) \right)}{\left(1-b^2 r_{+}^2\right) C_+}\,. 
\ee 
For the inner and outer black hole horizons, with the radius $r_-$ and $r_+$, one can find the area-temperature product in eq. (\ref{eq.areatempPROD}) is satisfied. Interestingly, the area temperature product presented in the last equation does not contain the boost parameter $\cosh\alpha$. In other words, this product is just the result for accelerating Kerr spacetime as presented in \cite{Podolsky:2021zwr}. Nevertheless, the product cannot be established for the acceleration horizons as the expression (\ref{eq.AreaTempProdgen}) becomes singular as one evaluates it at $r_h=  b^{-1}$. 

\section{Microscopic entropy from Kerr/CFT correspondence}\label{sec.KerrCFT}

In this section, we show the microscopic calculation for the black hole entropy at the extremal state following the Kerr/CFT correspondence prescription \cite{Guica:2008mu}. The Kerr/CFT holography for accelerating black hole in Einstein-Maxwell theory has been performed in \cite{Astorino:2016xiy,Siahaan:2018wvh}, whereas the holography discussion for non-AKK black hole was given in \cite{Azeyanagi:2008kb,Huang:2010yg,Li:2010ch}. To begin, we consider the following near-horizon of extremal state coordinate transformation
\be \label{eq.nearCOORD}
r \to a + \varepsilon r_0 y~~,~~t \to r_0 \frac{\tau }{\varepsilon }~~,~~\phi  \to \varphi + \frac{{r_0  C_{\rm ex.} }}{{2m\varepsilon \cosh \alpha }} \tau \,,
\ee 
with
\be 
r_0 = \frac{a \sqrt{2 \cosh \alpha}}{1-b^2 a^2}\,,
\ee 
and $C_{\rm ex.} = C_+ \left(m=a\right) = \left(1-ab\right)^2$. It yields the metric for near-horizon of extremal black hole geometry reads
\be \label{eq.nearmetric}
ds^2 = {\bar g}_{\mu\nu} dx^\mu dx^\nu  = \Gamma (x) \left( { - y^2 d\tau ^2  + \frac{{dy^2 }}{{y^2 }} + \gamma (x) dx^2} \right) + \beta \left(x\right)\left( {\frac{d\varphi }{C_{\rm ex.}} + {ky}d\tau } \right)^2 \,,
\ee
where 
\be 
k=\frac{1}{\left({1 - b^2 a^2 }\right)} \,,
\ee 
\be 
\Gamma (x)  = \frac{{a^2 \sqrt {\left( { 2\cosh ^2 \alpha -\Delta_x} \right)\left( {1 + x^2 } \right)} }}{{\left( {1 - bax} \right)^2 \left( {1 - b^2 a^2 } \right)}}\,,
\ee 
\be 
\gamma (x)  = \frac{{1 - b^2 a^2 }}{{\left( {1 - bax} \right)^2 \Delta_x}} \,,
\ee 
and
\be 
\beta \left( x \right) = \frac{{4a^2 \Delta _x \cosh ^2 \alpha }}{{\sqrt {\left( {1 + x^2 } \right)\left( {2\cosh ^2 \alpha  - \Delta _x } \right)} }}\,.
\ee 
In solving the equations of motion (\ref{eq.Einstein}) - (\ref{eq.Phi}), the Maxwell and dilaton fields can be found to be
\be 
{\bar A}_\mu  dx^\mu   = \frac{{a\Delta _x \sqrt {\cosh ^2 \alpha  - 1} \left( {kyd\tau + C_{\rm ex.}^{-1} d\varphi } \right)}}{{1 - x^2  - 2\cosh ^2 \alpha }}\,,
\ee 
and
\be 
{\bar \Phi}^2 = -\frac{1 - x^2  - 2\cosh ^2 \alpha}{1+x^2}\,,
\ee 
respectively. Note the warped and twisted $AdS_2 \times S^2$ structure of the near horizon geometry described by the metric (\ref{eq.nearmetric}). It strongly suggests the applicability of Kerr/CFT correspondence to reproduce the extremal black hole entropy in AKK spacetime.  

The prescription to compute the central charge associated to the conformal symmetry of the extremal Kaluza-Klein black hole has been given in \cite{Azeyanagi:2008kb,Li:2010ch}. The method can be repeated here which include the boundary conditions for diffeomorphisms of the fields, namely ${\cal L}_{\zeta_n} {\bar g}_{\mu\nu} = h_{\mu\nu}$, ${\cal L}_{\zeta_n} {\bar A}_{\mu}$, and ${\cal L}_{\zeta_n}{\bar \Phi}$ generated by the vector field 
\be 
{\zeta_n} = -e^{-in\varphi} \frac{\partial}{\partial \varphi}-  iny e^{-in\varphi} \frac{\partial}{\partial y}\,.
\ee  
The vector field above satisfies the Virasoro algebra
\be 
i \left[\zeta_m,\zeta_n\right] = \left(m-n\right) \zeta_{n+m}\,.
\ee 
Using Barnich-Brandt method, each diffeomorphism is related to a conserved charge $Q_\zeta$ given by an integral over a spacelike manifold $\partial M$. The Dirac brackets between this conserved charges can be shown the take the form
\be 
\left\{ {Q_\zeta  ,Q_\xi  } \right\} = Q_{\left[ {\zeta ,\xi } \right]}  - \int\limits_{\partial M} {{\bf K}_\zeta  } 
\ee 
where the last term is known as the central term. The integrand in this central term contains a two-form
\be 
{\bf K}_\zeta  \left[ {h,g} \right] = \frac{1}{{64\pi }}\varepsilon _{\alpha \beta \mu \nu } K_\zeta ^{\alpha \beta } dx^\mu   \wedge dx^\nu  \,,
\ee 
where
\be 
K_\zeta ^{\mu \nu }  = \zeta ^\nu  \nabla ^\mu  h - \zeta ^\nu  \nabla _\alpha  h^{\mu \alpha }  + \frac{h}{2}\nabla ^\nu  \zeta ^\mu   - h^{\nu \alpha } \nabla _\alpha  \zeta ^\mu   + \zeta _\alpha  \nabla ^\nu  h^{\mu \alpha }  - \left( {\mu  \to \nu } \right) \,,
\ee 
and $\epsilon_{0123} = \sqrt{-{\det} \left[{\bar g}_{\mu\nu}\right]}$. The covariant derivative and lowering/raising index are done by using the metric tensor for the near-horizon of extremal geometry (\ref{eq.nearmetric}). Explicitly, the calculation of the central term can be written as
\be 
\int\limits_{\partial M} {K_{\zeta _n } } \left[ {L_{\zeta _m } g,g} \right] =  - \frac{i}{{12}}c\left( {n^3  + \beta n} \right)\delta _{n, - m} \,.
\ee 
A general formula to compute the central charge which consists of the near-horizon of extremal metric functions as appeared in (\ref{eq.nearmetric}) has been given in \cite{Compere:2012jk}, namely
\be \label{eq.centralGen}
c = 3k\int\limits_{ - 1}^1 {dx\sqrt {\Gamma \left( x \right)\gamma \left( x \right)\beta \left( x \right)} } \,.
\ee
For the near-horizon of extremal AKK black hole as indicated in eq. (\ref{eq.nearmetric}), the central charge is found to be
\be \label{eq.central}
c = \frac{{12a^2 \cosh \alpha  }}{\left({1 - b^2 a^2 }\right)^2}\,.
\ee 
In the limit of non-accelerating case, i.e. $b=0$, this central charge reduces to the one for extremal Kaluza-Klein black hole as obtained in\footnote{Note that the boost parameter $v$ in \cite{Li:2010ch} is related to the $\alpha$ parameter used in this paper as $\cosh \alpha \sqrt{1-v^2}=1$.} \cite{Li:2010ch}. 

To complete the prescription of Kerr/CFT correspondence in reproducing the extremal black hole entropy from Cardy formula, one needs to compute the generalized temperature with respect to the Frolov-Thorne vacuum. It can be done by the following equation where one starts with the typical eigen-modes of energy with the frequency $\omega$ and angular momentum with a quantum magnetic number $h$ for the test scalar
\be 
\exp \left( { - i\omega t + i \frac{h \phi }{C_{\rm ex.}}} \right) = \exp \left( { - i\left( {\omega  - \frac{h}{{2m\cosh \alpha }}} \right)\frac{{r_0 \tau }}{\varepsilon } + i\frac{h \varphi }{C_{\rm ex.}} } \right)\,,
\ee 
where it is understood that the near-horizon coordinate transformation (\ref{eq.nearCOORD}) has been employed. Acknowledging the r.h.s. of last equation as \cite{Guica:2008mu} $e^{ - in_R \tau  + in_L \phi } $, we have
\be 
n_R  = \left( {\omega  - \frac{h}{{2a\cosh \alpha }}} \right)\frac{{r_0 }}{\varepsilon } ~~{\rm and}~~ n_L = \frac{h }{C_{\rm ex.}} \,.
\ee 
Then, an equation for the Boltzmann factor \cite{Compere:2012jk}
\be 
\exp \left( {  \frac{{\omega  - \Omega _H }}{{T_H }}} \right) = \exp \left( { - \frac{{n_L }}{{T_L }} - \frac{{n_R }}{{T_R }}} \right)
\ee 
yields the left and right temperature as
\be 
T_R  = -\frac{{r_0 T_H }}{\varepsilon }
\ee 
and
\be 
T_L =  \frac{1}{C_{\rm ex.}} \left[\mathop {\lim }\limits_{r_h  \to a}  \frac{{T_H }}{{\frac{1}{{2a\cosh \alpha }}-\Omega _H }}\right]  =
- \frac{1}{C_{\rm ex.}} \left. {\frac{{\partial T_H /\partial r_h }}{{\partial \Omega _H /\partial r_h }}} \right|_{r_h  = a} =\frac{1}{2\pi k ~C_{\rm ex.}} \,,
\ee 
respectively, where the angular velocity $\Omega_H$ is given in eq. (\ref{eq.OmH}). The Hawking temperature discussed above is the one corresponding to the outer black hole horizon, given by $r_h \to r_+$ in eq. \ref{eq.temp}. In the extremal limit, $a = m$, we have $T_H = 0$ which leads to $T_R =0$. For the left mode temperature, one can proceed as  
\be \label{eq.TL}
T_L = \frac{1}{2\pi k}\,.
\ee 
Note that the the extremal condition $T_H\to 0$ and near horizon limit $\varepsilon\to 0$ are two distinguished conditions. In obtaining $T_R =0$ above, $\varepsilon$ is not necessarily to be vanished \cite{Guica:2008mu}.

Finally, by using the Cardy formula
\be 
S_{CFT} = \frac{\pi^2}{3} c T_L
\ee 
with the central charge as given in eq. (\ref{eq.central}) and left temperature in (\ref{eq.TL}), one can obtain the entropy for an extremal AKK black hole
\be 
S_{{\rm{ext}}{\rm{.}}}  = \frac{{2\pi a^2 \cosh \alpha }}{{\left(1 - b^2 a^2 \right)}~C_{\rm ex.}}
\ee 
which is a quarter of black hole horizon area at extremal state in eq. (\ref{eq.area}). Therefore, it can be seen that Kerr/CFT correspondence can be used to show the black hole entropy at extremality in the case of AKK black hole.

\section{Conclusion}

In this study, we have derived an accelerated version of the Kaluza-Klein black hole solution and explored various aspects of this novel solution. Our examination of the electromagnetic field has uncovered discrepancies compared to the AKN scenario. Furthermore, we have investigated the area-temperature products for black hole horizons and observed that the expected relation holds true for both inner and outer horizons. Of particular interest is the discovery that the area-temperature product for black hole horizons in the AKK spacetime mirrors that of the seed solution, namely the accelerating Kerr spacetime. This intriguing result suggests a conjecture: in spacetimes obtained by transforming a seed solution belonging to the vacuum Einstein system, the area-temperature product remains consistent with that of the seed solution. The latter part of our study focused on verifying the generality of the Kerr/CFT correspondence, aiming to reproduce the extremal black hole entropy of the AKK case using the Cardy formula. 

As for future projects, we can outline several problems based on the findings of this paper. Firstly, we aim to validate the conjecture regarding the area-temperature product for black hole solutions obtained by transforming a vacuum Einstein system, comparing it with the relation holding in the seed system. This can be accomplished through the analysis of the accelerating Kerr-Sen solution \cite{Siahaan:2018qcw}, which arises from the Hassan-Sen transformation applied to the Kerr solution as the seed. Secondly, we plan to extend our investigation to a more general case involving the inclusion of the NUT parameter. Recent studies have reported an exact solution of the vacuum Einstein equations describing acceleration spacetime with the NUT parameter \cite{Podolsky:2021zwr,Podolsky:2020xkf}. It offers an opportunity to explore additional features in the AKK spacetime with the NUT parameter. 

Another possible interesting work to be pursued  is the investigation of the thermodynamics of the accelerating charged rotating spacetimes discussed in this paper. To the best of our knowledge, thermodynamic relations such as the Smarr formula and the first law have not been thoroughly explored for the accelerating Kerr-Newman spacetime \cite{Kim:2024dbj}. As demonstrated in this paper, the accelerating Kerr-Newman solution is simpler compared to the accelerating Kaluza-Klein case. Therefore, before studying the thermodynamics of the accelerating Kaluza-Klein spacetime, it is important to prioritize studies of these properties for the accelerating Kerr-Newman black hole. We plan to address this compelling problem in our future work.

\section*{Acknowledgement}

This project is supported by Kemdikbudristek.

\appendix

\section{Functions in the equatorial squared Riemann tensor}

The $c_i$'s functions in eq. (\ref{eq.RRx0}) are
\[
c_0= -3 \left( r-2m \right) ^{2} \left( 1-br \right) ^{2} \left( 1+br
\right) ^{2} \left( 5{m}^{4}{r}^{8}{b}^{8}+20{b}^{6}{m}^{4}{r}^{6
}-40{b}^{6}{m}^{3}{r}^{7}+10{b}^{6}{m}^{2}{r}^{8} \right.
\]
\[ +94{b}^{4}{m}^{
	4}{r}^{4}
-144{b}^{4}{m}^{3}{r}^{5}+116{b}^{4}{m}^{2}{r}^{6}-40{b
}^{4}{r}^{7}m+5{r}^{8}{b}^{4}-108{b}^{2}{r}^{2}{m}^{4}+88{b}^{2}
{r}^{3}{m}^{3}
\]
\[
\left. -22{b}^{2}{m}^{2}{r}^{4}+69{m}^{4}-64r{m}^{3}+16
{r}^{2}{m}^{2} \right) 
\]
\[
c_2 = 12 \left( r+2m \right)  \left( 1-b^2r^2 \right)  \left( 10{b}^{10}{r}^{10}{m}^{5}-5{b}^{10}{r}^{11}{m}^{4
}+30{b}^{8}{m}^{5}{r}^{8}-97{b}^{8}{m}^{4}{r}^{9}+60{b}^{8}{m}^{
	3}{r}^{10} \right.
\]
\[
+148{b}^{6}{m}^{5}{r}^{6}-306
{b}^{6}{m}^{4}{r}^{7}+388{b}^{6}{m}^{3}{r}^{8}-238{b}^{6}{m}^{2}{r
}^{9}+62{b}^{6}{r}^{10}m-404{b}^{4}{m}^{5}{r}^{4}+542{b}^{4}{m}^
{4}{r}^{5}
\]
\[
+202{b}^{4}{m}^{2}{r}^{7}-50{
	b}^{4}{r}^{8}m+354{b}^{2}{r}^{2}{m}^{5}-281{b}^{2}{m}^{4}{r}^{3}+
28{b}^{2}{m}^{3}{r}^{4}+6{b}^{2}{m}^{2}{r}^{5}+115r{m}^{4}
\]
\[
\left. -8{r
}^{2}{m}^{3}-8{r}^{3}{m}^{2} -404{b}^{4}{m}^{3}{r}^{6}-138{m}^{5}+4{b}^{2}m{r}^{6}+3{r}^
{9}{b}^{4}-5{r}^{11}{b}^{6} -10{b}^{8}{m}^{2}{r}^{11} \right) 
\]
\[
c_4 = 1144m{r}^{9}{b}^{4}-2016m{r}^{11}{b}^{6}+39928{m}^{4}{r}^{8}{b}^
{6}+10112{m}^{3}{r}^{7}{b}^{4}+2272{m}^{3}{r}^{5}{b}^{2}-5032{m}
^{2}{r}^{8}{b}^{4}
\]
\[
+3672r{m}^{5}+286{r}^
{2}{m}^{4}-384{r}^{3}{m}^{3}-27288{m}^{6}{b}^{4}{r}^{4}+17712{m}
^{6}{b}^{2}{r}^{2}-28160{m}^{3}{r}^{9}{b}^{6}+11144{m}^{2}{r}^{10}
{b}^{6}
\]
\[
-1772{m}^{4}{r}^{4}{b}^{2}-36864
{r}^{7}{m}^{5}{b}^{6}+31608{r}^{5}{m}^{5}{b}^{4}-12960{r}^{3}{m}
^{5}{b}^{2}+17280{b}^{8}{r}^{11}{m}^{3}-7704{b}^{8}{r}^{12}{m}^{2}
\]
\[
+1440{b}^{10}{r}^{13}{m}^{3}-180{b}^{10}{r}
^{14}{m}^{2}-360{m}^{6}{b}^{12}{r}^{12}+3744{m}^{5}{b}^{10}{r}^{11
}+10440{m}^{5}{b}^{8}{r}^{9}-3852{m}^{4}{b}^{10}{r}^{12}
\]
\[
+360{m}^{5}{b}^{12}{r}^{13}-90{m}^{4}{b}^{12}{
	r}^{14}-720{m}^{6}{b}^{10}{r}^{10}-4248{m}^{6}{b}^{8}{r}^{8}+19872
{m}^{6}{b}^{6}{r}^{6}-90{b}^{8}{r}^{14}+108{b}^{6}{r}^{12}
\]
\[
-48{m}^{2}{r}^{4}-4968{m}^{6}-17142{m}^{4}{r}^{6}{b}^{4}-18638{m
}^{4}{b}^{8}{r}^{10}-100{m}^{2}{r}^{6}{b}^{2}+1512{b}^{8}{r}^{13}m-98{
b}^{4}{r}^{10}
\]
\[
c_6 =
3968{m}^{2}{r}^{8}{b}^{4} -48{b}^{2}{r}^{7}m-1136m{r}^{9}{b}^{4}+864m{r}^{11}{b}^{6}-23456
{m}^{4}{r}^{8}{b}^{6}-5728{m}^{3}{r}^{7}{b}^{4}-1184{m}^{3}{r}^{
	5}{b}^{2}
\]
\[
+104{m}^{2}{r}^{6}{b}^{2}-480
r{m}^{5}-620{r}^{2}{m}^{4}-96{r}^{3}{m}^{3}+18192{m}^{6}{b}^{4}{
	r}^{4}-11808{m}^{6}{b}^{2}{r}^{2}+15712{m}^{3}{r}^{9}{b}^{6}
\]
\[
+4980{m}^{4}{r}^{6}{b}^{4}+4480{m}^{4}{r}^
{4}{b}^{2}+22176{r}^{7}{m}^{5}{b}^{6}-13296{r}^{5}{m}^{5}{b}^{4}+
1872{r}^{3}{m}^{5}{b}^{2}-13632{b}^{8}{r}^{11}{m}^{3}
\]
\[
-1152{b}^{8}{r}^{13}m-960{b}^{10}{r}^{13}{m}^{3}+
120{b}^{10}{r}^{14}{m}^{2}+240{m}^{6}{b}^{12}{r}^{12}-2544{m}^{5}{b}^{10}{r}^{11}-7488{m}^{5}{b}^{8}{r}^{9}
\]
\[
+14524{m}^{4}{b}^{8}{r}^{10}-240{m}^{5}{b}^{12}{r}^{13}+60{m}^{4}{b}^{12}{r}^{14}+480{m}^{6}{b}^{10}{r}^{10}+2832{m}^{6}{b}^{8}{r}^{8}-13248{m}^{6}{b}^{6}{r}^{6}+60{b}^{8}{r}^{14}
\]
\[
+148{b}^{4}{r}^{10}+3312{m}^{6}-5632
{m}^{2}{r}^{10}{b}^{6}-48{b}^{6}
{r}^{12}+2592{m}^{4}{b}^{10}{r}
^{12}+6048{b}^{8}{r}^{12}{m}^{2}
\]
and
\[
c_8 = 532m{r}^{9}{b}^{4}-24m{r}^{11}{b}^{6}+4924{m}^{4}{r}^{8}{b}^{6}+
1760{m}^{3}{r}^{7}{b}^{4}+112{m}^{3}{r}^{5}{b}^{2}-1516{m}^{2}{r}^{8}{b}^{4}-238{m}^{2}{r}^{6}{b}^{2}
\]
\[
-71{r}^{2}{m}^{4}-4548{m}^{6}{b}^{4}{r}^{4}+2952{m}^{6}{b}^{2}{r}^{2}-2720{m}^{3}{r}^{9}{b}^{6}+644{m}^{2}{r}^{10}{b}^{6}-225{m}^{4}{r}^{6}{b}^{4}-1070{m}^{4}{r}^{4}{b}^{2}
\]
\[
+1380{r}^{5
}{m}^{5}{b}^{4}+1224{r}^{3}{m}^{5}{b}^{2}+3936{b}^{8}{r}^{11}{m}^{3}-1740{b}^{8}{r}^{12}{m}^{2}+324{b}^{8}{r}^{13}m+240{b}^{10}{r}
^{13}{m}^{3}-30{b}^{10}{r}^{14}{m}^{2}
\]
\[
+
648{m}^{5}{b}^{10}{r}^{11}+2004{m}^{5}{b}^{8}{r}^{9}-654{m}^{4}{b}^{10}{r}^{12}-4169{m}^{4}{b}^{8}{r}^{10}+60{m}^{5}{b}^{12}{r}^{
	13}-15{m}^{4}{b}^{12}{r}^{14}-120{m}^{6}{b}^{10}{r}^{10}\]
\[
+3312{m}^{6}{b}^{6}{r}^{6}-15{b}^{8}{r}^{14}+6{b}^{6}{r}^{12}-71{b}^{4}{r}^{10}-828{m}^{6}-60{m}^{6}{b}^{12}{r}^{12}-708{m}^
{6}{b}^{8}{r}^{8}-372r{m}^{5}-4944{r}^{7}{m}^{5}{b}^{6}
\]

\section{Central charge and a scaling on $\varphi$ coordinate}

Under the coordinate scaling $\varphi \to  C_{\rm ex.}^{-1}~\varphi$, the near horizon metric (\ref{eq.nearmetric}) transforms to
\be \label{eq.nearmetricNEW}
ds^2  = \Gamma \left( { - y^2 d\tau  + \frac{{dy^2 }}{{y^2 }} + \gamma dx^2 } \right) + \beta ' \left( {d\varphi  + k' yd\tau } \right)^2 
\ee 
where $\beta ' = C_{\rm ex.}^{-2} \beta $ and $k'=k C_{\rm ex.}$. Obviously, by using the general formula (\ref{eq.centralGen}), the central charge associated to the new near horizon metric (\ref{eq.nearmetricNEW}) can be written as
\be\label{eq.central2} 
c'  = 3k' \int\limits_{ - 1}^1 {dx\sqrt {\Gamma \gamma \beta ' } } 
\ee 
which turns out to be equal with the seed central charge, i.e. $c ' =c$. Here we understand the invariance of central charge under a scale in $\varphi$ coordinate.

\end{document}